6/29/12

# The Classical Electromagnetic Modes of a Rectangular Resonant Cavity with particular reference to the text Quantum Physics by R. Eisberg and R. Resnick

W. Zimmermann, Jr.

Tate Laboratory of Physics, University of Minnesota
Minneapolis, Minnesota, U.S.A. 55455

## Abstract

Three approaches to the derivation of the classical electromagnetic modes of a rectangular cavity are described. In so doing, some apparent errors in a widely used physics text are pointed out.

Some time ago I was asked by a colleague to give a lecture to his sophomore physics class on thermal cavity radiation. In the course of trying to understand some puzzling statements in the first chapter of the widely-used text by Eisberg and Resnick, entitled Quantum Physics, about accounting for the normal modes of a cavity, I came to realize that there were several equivalent approaches to the subject giving different insights. Because I am not aware of any reference that presents these various approaches in one place, I thought that it might be useful to summarize and relate them. This note also gives me the opportunity to point out some statements in Eisberg and Resnick that I believe to be in error, as well as one or two misleading statements in other places. [1]

### A. Direct Approach

Let's assume that we are dealing with an empty rectangular cavity with perfectly conducting walls, with edges of length $a_x$, $a_y$, and $a_z$ lying parallel to the $x$, $y$, and $z$ axes, respectively, and that lies in the first octant of the coordinate system with one corner at the origin. If then we look for normal-mode solutions of Maxwell's equations, that is, standing wave solutions in which all of the field components vary sinusoidally at a single frequency, that satisfy the boundary conditions, namely, that the tangential components of the electric field $E$ and the normal components of the magnetic induction $B$ must vanish at the walls, then we can find the following expressions for the components of the fields $E$ and $B$: [2-4]

$$E_x = E_{x0}\cos(k_x x)\sin(k_y y)\sin(k_z z)e^{-i\omega t} \qquad B_x = B_{x0}\sin(k_x x)\cos(k_y y)\cos(k_z z)e^{-i\omega t}$$

$$E_y = E_{y0}\sin(k_x x)\cos(k_y y)\sin(k_z z)e^{-i\omega t} \qquad B_y = B_{y0}\cos(k_x x)\sin(k_y y)\cos(k_z z)e^{-i\omega t}$$

$$E_z = E_{z0}\sin(k_x x)\sin(k_y y)\cos(k_z z)e^{-i\omega t} \qquad B_z = B_{z0}\cos(k_x x)\cos(k_y y)\sin(k_z z)e^{-i\omega t}.$$

(1)

Here, $k_x = n_x \pi / a_x$, $k_y = n_y \pi / a_y$, and $k_z = n_z \pi / a_z$, where $n_x$, $n_y$, and $n_z$ are independent non-negative integers subject only to the restriction that no more than one can be zero at a time. The angular frequency of oscillation $\omega$ is equal to $kc$, where $k$ is the magnitude of the vector **k** whose components are $k_x$, $k_y$, and $k_z$, and $c$ is the velocity of light. Thus $\omega$ is given by

$$\omega = (n_x^2 \pi^2 / a_x^2 + n_y^2 \pi^2 / a_y^2 + n_z^2 \pi^2 / a_z^2)^{1/2} c. \tag{2}$$

For convenience here and in what follows, we use complex forms with the understanding that the real parts may be taken as the physically meaningful quantities. Phase information is then carried by the complex amplitudes $E_{x0}$, $E_{y0}$, $E_{z0}$, $B_{x0}$, $B_{y0}$, and $B_{z0}$.

Perhaps the simplest way to derive the expressions for the components of **E** is to note that the components individually satisfy identical wave equations. Solutions to these equations by the method of separation of variables together with the boundary conditions and the condition *div* **E** = *0* then lead to the expressions above for the components of **E**. [2-4] The three amplitudes $E_{x0}$, $E_{y0}$, and $E_{z0}$ are not independent of one another, for the condition *div* **E** = *0* requires that for each set of *n*-values the amplitudes $E_{x0}$, $E_{y0}$, and $E_{z0}$ form a vector that is perpendicular to **k**. But note that it is not correct to conclude, as is sometimes stated, that **E** itself is perpendicular to **k** for a given set of *n*-values, since for many modes, the direction of **E** varies within the cavity. [2,3] An exception to this remark exists when one of the *n*-values is zero, for then the alignment of **E** is unique and perpendicular to **k**.

The expressions for the components of **B** follow from those for **E** and the Maxwell equation *curl* **E** = -∂**B**/∂*t* (SI units are being used throughout this article.). The amplitudes $B_{x0}$, $B_{y0}$, and $B_{z0}$ are given in terms of $E_{x0}$, $E_{y0}$, and $E_{z0}$ by the following expressions:

$$B_{x0} = \frac{k_y E_{z0} - k_z E_{y0}}{i\omega} \qquad B_{y0} = \frac{k_z E_{x0} - k_x E_{z0}}{i\omega} \qquad B_{z0} = \frac{k_x E_{y0} - k_y E_{x0}}{i\omega}. \tag{3}$$

These amplitudes thus form a vector $\mathbf{B}_0 = \mathbf{k} \times \mathbf{E}_0/i\omega$, perpendicular to both $\mathbf{E}_0$ and **k**. As with **E**, **B** itself is not in general perpendicular to **k**. However, **B** is everywhere perpendicular to **E**. The imaginary $i$ in these expressions means e.g. that if the variations of the components of the **E** field are all in phase with each other, variations of all of the components of the **B** field are 90 degrees out of phase with them. Since for each **k** for which $n_x$, $n_y$, and $n_z$ are all non-zero there are two linearly-independent directions for $\mathbf{E}_0$ (or for its corresponding $\mathbf{B}_0$), we have two modes having the same value of $\omega$ for each set of *n*-values, i.e. two polarizations. It is conventional to choose one to have $E_{z0} = 0$, the "transverse electric (TE) mode", and one to have $B_{z0} = 0$, the "transverse magnetic (TM) mode". When one of the *n*-values in a set is zero there is only one mode for the set. When $n_x$ or $n_y$ is zero, the mode is transverse electric. When $n_z$ is zero, the mode is transverse magnetic.

The first treatment of these normal modes given by Eisberg and Resnick on page 7 of Quantum Physics appears to be erroneous. [1] It is based on the assumption that the modes can be considered to be superpositions of independent non-interfering standing waves along each of the axis directions. This assumption does not seem reconcilable with the expressions for the components of *E* and *B* given above.

## B. Waveguide Approach

A second approach to this problem is first to consider the traveling-wave modes of a waveguide of rectangular cross-section with sides $a_x$ and $a_y$ parallel to the *x* and *y* axes, respectively, and then to form from these modes the standing waves that result from blocking the two ends of a section of the waveguide of length $a_z$ along the *z*-axis with conducting walls parallel to the *x,y* plane. [5,6] The traveling wave modes are of two types, transverse electric and transverse magnetic. In the transverse electric (TE) case, the *E* fields are perpendicular to the *z* axis, the direction of propagation, whereas in the transverse magnetic (TM) case, the *B* fields are perpendicular to the direction of propagation. Although it is usual to treat these two cases separately, [3-10] it is convenient to begin by considering them together. [11]

Expressions for *E* and *B* can be written:

$$E_x = -iE_{x0}'\cos(k_x x)\sin(k_y y)\exp[i(k_z z - \omega t)] \qquad B_x = B_{x0}'\sin(k_x x)\cos(k_y y)\exp[i(k_z z - \omega t)]$$
$$E_y = -iE_{y0}'\sin(k_x x)\cos(k_y y)\exp[i(k_z z - \omega t)] \qquad B_y = B_{y0}'\cos(k_x x)\sin(k_y y)\exp[i(k_z z - \omega t)]$$
$$E_z = E_{z0}'\sin(k_x x)\sin(k_y y)\exp[i(k_z z - \omega t)] \qquad B_z = -iB_{z0}'\cos(k_x x)\cos(k_y y)\exp[i(k_z z - \omega t)],$$
(4)

where

$$B_{x0}' = \frac{k_y E_{z0}' - k_z E_{y0}'}{i\omega} \qquad B_{y0}' = \frac{k_z E_{x0}' - k_x E_{z0}'}{i\omega} \qquad B_{z0}' = \frac{k_x E_{y0}' - k_y E_{x0}'}{i\omega}.$$
(5)

Here $k_x = n_x \pi /a_x$ and $k_y = n_y \pi /a_y$, where $n_x$ and $n_y$ are non-negative integers, as above, at least one of which must be non-zero, (save for the trivial solution in which *E* = *0* and *B* = *zB_z*, uniform and constant) but $k_z$ is unconstrained save for our requirement that it be non-negative. As above, $\omega = kc$. The factors *-i* have been introduced to make the phases of the amplitudes compatible with each other and with the choices made in Section A. As a result the conditions *div E* = *0* and *div B* = *0* yield *E₀´* and *B₀´* vectors that are perpendicular to *k*, as in Section A, and the relations Eqs. (5) obeyed by their components are of the same form as Eqs. (3). Thus *E₀´* and *B₀´* satisfy the same relationship as *E₀* and *B₀* and are mutually perpendicular.

These results may be derived using the same approach as outlined above for the cavity modes, except that here the boundary conditions at *z* = *0* and *z* = $a_z$ are absent and traveling waves in the positive *z*-direction are sought. [9] When $k_x$ and $k_y$ are both non-zero, there exist both TE solutions (with $E_z = 0$) and TM solutions (with $B_z = 0$) having the same value of ω for each such set of k-values, for all values of $k_z$ including $k_z = 0$, even though for $k_z = 0$ the modes are non-propagating modes. When $k_x$ and $k_y$ are both non-zero, the most general solution can be written as a linear combination of the TE and TM solutions. When either $k_x$ or $k_y$ is zero, only TE solutions exist, for all values of $k_z$ including $k_z = 0$.

Now let us block the waveguide at $z = 0$ and $z = a_z$ with plane, perfectly-conducting partitions. In order to satisfy the boundary conditions *E(tangential)* = 0 and *B(normal)* = 0 at the partitions, we must superimpose a reflected wave of the same amplitude, having the same $k_x$ and $k_y$ but traveling in the opposite direction along the z-axis to the original wave so as to produce a standing wave with components varying with z as $sin(k_z z)$ or $cos(k_z z)$, with $k_z = n_z \pi / a_z$. In the TE case, $n_z$ must be greater than zero; the non-propagating TE wave-guide modes cannot satisfy the boundary conditions at the partitions and therefore cannot form cavity modes. For the TM case, $n_z = 0$ must be retained as a possibility.

The result is a set of standing waves having just the same set of spatial dependences as in Eq. (1). The amplitude coefficients are given by

$$E_{x0} = 2E_{x0}' \qquad E_{y0} = 2E_{y0}' \qquad E_{z0} = 2E_{z0}'$$
$$B_{x0} = 2B_{x0}' \qquad B_{y0} = 2B_{y0}' \qquad B_{z0} = 2B_{z0}', \qquad (6)$$

and the relationships between the *E* and *B* coefficients are the same as in Eqs. (3). The degeneracies are the same, and the modes derived from TE and TM traveling waves respectively are just the TE and TM cavity modes defined earlier. So in the end, this second approach yields just the same spectrum of modes as the first approach.

## C. Free Traveling-Wave Approach

A third approach to this problem is to start with a plane wave having an arbitrary polarization traveling in an initially arbitrary direction within the cavity of the form:

$$\boldsymbol{E} = \boldsymbol{E}_0'' exp(i\boldsymbol{k}^*\boldsymbol{r} - \omega t) \qquad \boldsymbol{B} = \boldsymbol{B}_0'' exp(i\boldsymbol{k}^*\boldsymbol{r} - \omega t), \qquad (7)$$

where $\boldsymbol{E}_0'' \cdot \boldsymbol{k} = 0$ and $\boldsymbol{B}_0'' = \boldsymbol{k} \times \boldsymbol{E}_0''/\omega$. Each time this wave encounters a wall of the cavity it will be specularly reflected with a reversal of its motion along the axis perpendicular to the wall, a reversal of the electric field components parallel to the wall, and a reversal of the magnetic field component perpendicular to the wall. When we take into account all such reflections, together with reflections of reflections and the original wave, we end up with eight plane waves of equal amplitude having *k*-vectors made up of the eight possible combinations of ± $k_x$, ±$k_y$, and ±$k_z$. Each of these waves has a polarization that follows uniquely from the polarization of the original wave. The requirement that the contributions to any one of the eight waves must interfere constructively limits the values of $k_x$, $k_y$, and $k_z$ to the same set found for the standing waves in Sections A and B above. As a result, when the eight traveling waves are superimposed, they yield the standing waves of Eqs. (1) with

$$\boldsymbol{E}_0 = -8\boldsymbol{E}_0'' \qquad \boldsymbol{B}_0 = 8i\boldsymbol{B}_0''. \qquad (8)$$

I was tempted at first to think of this approach in terms of two corner reflectors, each with surfaces parallel to those of the other, facing each other. It would seem in this case that only three reflections taking place in each corner reflector would suffice to restore an initial beam of radiation to its original direction and polarization, although perhaps not to a coincident path. Only a total of six propagation directions would be involved. However, in the present case, in which the waves are non-local, it seems essential to consider all possible reflections together.

Perhaps Eisberg and Resnick had in mind an approach similar to this third approach in their second discussion of cavity modes given in Example 1-3 beginning on page 10 of their text. [1] However, if by nodal planes they mean planes on which the $\boldsymbol{E}$ field or perhaps the $\boldsymbol{B}$ field vanishes, their notion that there exist such planes perpendicular to an arbitrary single propagation axis allowed by the boundary conditions seems erroneous, at least for the normal modes given by Eqs. (1). As described above, each standing wave generated by a set of traveling waves involves in general four different propagation axes. Moreover, the Cartesian components of the electric field and the magnetic induction individually have nodal planes that are perpendicular to the coordinate axes for the modes described by Eqs. (1). So for those modes the magnitude of either field could vanish only on planes of this sort.

The standing waves portrayed by Eisberg and Resnick seem to correspond to superpositions of two counter-propagating traveling waves along a given axis of propagation. But such standing waves do not satisfy the boundary conditions.

It must be noted that when degeneracies of modes of the same polarization occur, linear superpositions of such modes can yield nodal surfaces for the individual field components other than those mentioned above, nodal surfaces that can have quite fanciful shapes. The existence of such degeneracies depends on the relations between the lengths $a_x$, $a_y$, and $a_z$, and, indeed, the cubical case treated by Eisberg and Resnick is particularly rich in degeneracies. However, their treatment seems directed toward simpler and more general normal mode solutions, of the sort presented above, than those involving superpostions of such solutions.

It is interesting to note that the free traveling-wave approach described above can also be used to generate the TE and TM modes of rectangular waveguides. [6] In general for these cases, the superposition of four traveling plane waves having specific polarization axes is required. In the framework used above, for a given set of values for $k_x$, $k_y$, and $k_z$, these waves would be characterized by the four combinations of $\pm k_x$ and $\pm k_y$ with $k_z$.

In conclusion, it is gratifying to see how all three of the approaches to the derivation of the electromagnetic modes of a rectangular cavity described here, with their various insights, yield identical results. I am indebted to my colleague Professor B. Bayman for pointing out to me the possible role of degeneracies in regard to the second discussion of Eisberg and Resnick.